\documentclass[11pt,a4paper]{article}
\usepackage{pos}
\usepackage{subcaption}
\usepackage{caption}

\usepackage{mathtools}
\DeclareMathOperator{\ReN}{Re}

\DeclareMathOperator{\tr}{Tr}
\newcommand{\dif}[1]{\mathrm{d} #1 }
\newcommand{\effAction}{S^{\text{eff}}}
\newcommand{\staticEffAction}{\effAction_{\text{S}}}
\newcommand{\nonStaticEffAction}{\effAction_{\text{G,K}}}
\newcommand{\pureGaugeEffAction}{\effAction_{\text{G}}}
\newcommand{\kineticQuarkEffAction}{\effAction_{\text{K}}}

\title{Finite density lattice QCD via effective Polyakov loop theories}
\author[a]{Christoph Konrad}
\author[a]{Owe Philipsen}
\affiliation[a]{Institute for Theoretical Physics, Goethe-University Frankfurt am Main, Max-von-Laue-Str. 1, 60438 Frankfurt am Main, Germany}
\emailAdd{konrad@itp.uni-frankfurt.de}
\emailAdd{philipsen@itp.uni-frankfurt.de}

\abstract{For the exploration of the phase diagram of QCD, effective Polyakov loop theories derived from lattice QCD provide a valuable tool in the heavy quark mass regime. Using mean field approximations these theories are evaluated in the high and low temperature regimes at finite baryon chemical potential. The resulting phase diagram is discussed.}

\FullConference{The 41st International Symposium on Lattice Field Theory (LATTICE2024)\\
	Liverpool, UK\\
	28 July - 3 August 2024
}

\begin{document}
	\maketitle
	
\section{Introduction}
The determination of the QCD phase diagram is hindered by the sign problem in Monte Carlo evaluations of lattice QCD (LQCD) at non-zero baryon chemical potential $\mu_B$. To address this, effective theories have been derived from Wilson's standard lattice action with arbitrary $\mu_B$ using strong coupling and hopping parameter expansions~\cite{Langelage2011, Langelage:2014vpa, Glesaaen2016}. These effective theories, formulated in three dimensions with Polyakov loops as field variables, are computationally less demanding and exhibit a significantly reduced sign problem compared to the mother theory. This enables their numerical evaluation at $\mu_B \neq 0$~\cite{Fromm2012, Fromm2013, Langelage:2014vpa, Glesaaen2016}. Additionally, they allow the application of series expansion techniques that are not sensitive to the sign problem~\cite{Langelage:2014vpa, Glesaaen2016, Philipsen:2019qqm}. 

These effective theories were found to describe the critical couplings for SU(3) pure gauge theory within 10\% accuracy~\cite{Langelage2011}, and the second-order critical deconfinement temperature line was mapped out for varying heavy quarks masses and $\mu_B$~\cite{Fromm2012}. At low temperatures earlier studies examined the baryon onset transition \cite{Langelage:2014vpa, Fromm2013} and identified a quarkyonic regime for large number of colors $N_c$~\cite{Philipsen:2019qqm}. However, higher-order corrections to the expansions, which are necessary for describing lighter quarks and finer lattices, introduce long-range interactions.

Mean field approximations, whose accuracy is well known to increase if interactions become long-range, have been applied successfully to Polyakov loop models~\cite{Dumitru:2005ng, Fukushima:2006uv, Greensite:2012xv, Rindlisbacher:2015pea, Borisenko:2020cjx}. This work refines these studies by introducing a resummation scheme for local fluctuations, leading to an improved accuracy for first order transitions, and applying mean field to effective theories, which describe LQCD to higher orders in the expansion parameters.

We apply three mean field variants and compare their predictions for the deconfinement transition against earlier results at zero and non-zero $\mu_B$. For first order transitions at high temperatures the naive mean field approach shows $\approx 50$\% error, while the most involved variant improves the accuracy to $\approx 3$\%. All approaches predict a second-order critical endpoint, though its location is imprecise due to fluctuation-driven dynamics. At low temperatures, we confirm the appearance of a first-order liquid-gas transition for light quarks, in agreement with complex Langevin simulations~\cite{Langelage:2014vpa}. The behavior of the entropy density across this transition differs from expectations for the continuum theory, 
but is found to be consistent with earlier perturbative analyses~\cite{Langelage:2014vpa}.

\section{Overview of effective theories \label{sec:Overview_effective_theories}}

Effective theories are derived from the lattice QCD after integrating out $N_f$ mass-degenerate quarks with identical chemical potentials \cite{Langelage2011, Fromm2012},
\begin{align}
	Z &= \int[\dif{U_\mu}] e^{-S_G[U_\mu]} \left(\det Q[U_\mu]\right)^{N_f},
\end{align}
where $S_G$ is the Wilson gauge action, and $Q$ is the Wilson-Dirac operator. The quark determinant splits into static and kinetic contributions \cite{Langelage:2014vpa},	$\det Q = \det Q_{\text{S}} \det Q_{\text{K}}$,
where $\det Q_{\text{S}}$ describes purely temporal quark hops and factorizes over spatial lattice sites. The kinetic quark determinant $\det Q_{\text{K}}$ accounts for spatial quark hops. The effective action $\effAction$, which splits into the static effective action $\staticEffAction$ and non-static effective action $\nonStaticEffAction$ with $\effAction := \staticEffAction + \nonStaticEffAction$, is defined after spatial links have been integrated out,
\begin{align}
	Z &= \int{\dif{U_0}} \exp(-\effAction),\quad 
	-\staticEffAction := N_f \ln \det Q_\text{S}, \quad
	-\nonStaticEffAction := \ln \int[\dif{U_{\upsilon\neq 0}}] e^{-S_G} (\det Q_{\text{K}})^{N_f}.
\end{align}
 $\effAction$ is a functional of traces of temporal Wilson lines $W_{\mathbf{x}} := \prod_{\tau=0}^{N_\tau-1} U_0(\mathbf{x}, \tau)$ because of gauge invariance. In practice, approximations are necessary to derive expressions for $\nonStaticEffAction$, which motivates the application of combined strong-coupling and hopping parameter expansions~\cite{Langelage2011}. 

The pure gauge effective action $\pureGaugeEffAction$ has been derived via the application of character expansions and the linked cluster theorem. The leading contribution is a nearest-neighbor interaction~\cite{Langelage2011},
\begin{align}
	-\pureGaugeEffAction &= \sum_{\langle \mathbf{x}, \mathbf{y} \rangle} 
	\ln\left(1 + \lambda_1 L_{\mathbf{x}} L_{\mathbf{y}}^* + \lambda_1 L_{\mathbf{y}} L_{\mathbf{x}}^*\right) 
	+ \mathcal{O}(\beta^{2N_\tau}), \label{eq:pure_gauge_effective_action}
\end{align}
where $L_{\mathbf{x}} := \tr W_{\mathbf{x}}$. The expression for the effective coupling $\lambda_1 = \lambda_1(\beta, N_\tau)$ used within this work is given in \cite{Langelage2011}. Higher-order terms introduce long-range interactions and interactions between higher representations of Polyakov loops~\cite{Langelage2011}. These are neglected within this work.

In the strong coupling limit the kinetic quark effective action $S^{\text{eff}}_{\text{K}}$ is derived by expanding $\det Q_{\text{K}}$ using $\det(\cdot) = \exp(\tr \ln(\cdot))$. After spatial link integration $S^{\text{eff}}_{\text{K}}$ may be expressed in terms of
\begin{align}
	W_{nm\bar{n}\bar{m}}(W_{\mathbf{x}}) &= \tr \left(\frac{(h_1 W_{\mathbf{x}})^m}{(1 + h_1 W_{\mathbf{x}})^n} 
	\frac{(\bar{h}_1 W_{\mathbf{x}}^{-1})^{\bar{m}}}{(1 + \bar{h}_1 W_{\mathbf{x}}^{-1})^{\bar{n}}}\right), \\
	h_1 &= (2\kappa)^{N_\tau}e^{N_\tau a\mu} (1+\dots),\quad\bar{h}_1 = h_1(-\mu). \label{eq:Definition_h1}
\end{align}
For the physically interesting case $N_c = 3$ the $W_{nm\bar{n}\bar{m}}$ can always be written as functions of the  Polyakov loops $L$ and $L^*$. This can be achieved via generating function techniques \cite{Glesaaen2016} or the Cayley-Hamilton theorem \cite{Rindlisbacher:2015pea}. The leading-order kinetic quark effective action is
\begin{align}
	-\kineticQuarkEffAction = -2N_f h_2 \sum_{\langle \mathbf{x}, \mathbf{y} \rangle} 
	W_{1111}^-(W_{\mathbf{x}}) W_{1111}^-(W_{\mathbf{y}}) + \mathcal{O}(\kappa^4),\quad h_2 = \frac{N_\tau \kappa^2}{N_c}(1 + \dots). \label{eq:LO_kinetic_quark_eff_action}
\end{align}
Higher-order corrections introduce non-local interactions~\cite{Langelage:2014vpa, Glesaaen2016}.

Beyond leading order the effective couplings depend on all LQCD parameters \cite{Fromm2012, Langelage:2014vpa}. 
Here, $\nonStaticEffAction$ is approximated as $\nonStaticEffAction(\beta, \kappa, N_\tau) \approx 
	\kineticQuarkEffAction(\beta, \kappa, N_\tau) + \pureGaugeEffAction(\beta, \kappa, N_\tau)$.
In this work we consider effective actions correct to $\mathcal{O}(u^n \kappa^m)$ with $n+m \leq 4$, as given in \cite{Langelage:2014vpa}, with gauge corrections to the kinetic quark effective couplings as given in sections 4.4.1 and 4.4.3 of~\cite{NeumanPhD}.

\section{Mean Field Approximations \label{sec:General_mean_field_discussion}}
Within mean field approximations the action is expressed through self-consistent mean fields $l$ and $\bar{l}$ and fluctuations, $\delta L_{\mathbf{x}} := L_{\mathbf{x}} - l$ and $\delta L_{\mathbf{x}}^* := L_{\mathbf{x}}^* - \bar{l}$, around them. The action is then Taylor-expanded in the fluctuations. We demonstrate this by considering an expression with a formally general nearest-neighbor effective Polyakov loop interaction $I(L_{\mathbf{x}}, L_{\mathbf{x}}^*, L_{\mathbf{y}}, L_{\mathbf{y}}^*)$,
\begin{align}
	Z = \int [\dif{W_{\mathbf{x}}}] \exp\left(\sum_{\mathbf{x}} 
	\ln \det Q_\text{stat}^\text{loc}(L_{\mathbf{x}}, L_{\mathbf{x}}^*) + 
	\sum_{\langle \mathbf{x}, \mathbf{y} \rangle} I(L_{\mathbf{x}}, L_{\mathbf{x}}^*, L_{\mathbf{y}}, L_{\mathbf{y}}^*)\right), \label{eq:pgStatic}
\end{align}
We now discuss three types of mean field approximations which capture different levels of local fluctuations and complexity of the original system \eqref{eq:pgStatic}.
		\subsection{Standard mean field approximation\label{sec:smf}}
After expanding the interaction $I$ to $\mathcal{O}(\delta L)$  
the effective action is approximated by purely local interactions between the Polyakov loops and the mean fields. The partition function factorizes \cite{Zinn-Justin202}, i.e. $Z \approx z_{\text{s-mf}}^V$ where
	\begin{align}
		z_{\text{s-mf}} = \int\dif{W}\det Q^{\text{loc}}_{\text{stat}}(W)\exp\left[dI(l, \bar{l}, l, \bar{l}) +d\left( (L - l) \frac{\partial}{\partial l} + (L^* - \bar{l}) \frac{\partial}{\partial \bar{l}}\right)I(l, \bar{l}, l, \bar{l})\right] \;,\label{eq:s-mf_partitionfunction}
	\end{align}
	with the number of spatial dimensions $d$ and the number of lattice sites $V$. To approximate $Z$ one then computes the single-site integral \eqref{eq:s-mf_partitionfunction} and solves the two coupled self-consistency equations
		$l = \langle L \rangle_{\text{s-mf}}(l,\bar{l}) \quad\text{and}\quad \bar{l} = \langle L^* \rangle_{\text{s-mf}}(l,\bar{l})$.
	From these solutions the one with the largest $z_{\text{s-mf}}$ represents the best approximation to the true partition function $Z$.	Instead of directly determining the self-consistent $l,\bar{l}$ one may locate the saddle points of the free energy density $a^4f_{\text{s-mf}} = -\ln (z_{\text{s-mf}})/N_\tau$.
	The saddle point with the lowest $f_{\text{s-mf}}$ corresponds to the physically relevant mean fields \cite{Zinn-Justin202}. 
	\subsection{Resummed mean field approximation \label{sec:rmf}}
  To improve the mean field approximation's accuracy, we now introduce an approximation scheme that resums a subset of fluctuations to all orders. As in section~\ref{sec:smf}, we neglect non-local terms $\mathcal{O}(\delta L_{\mathbf{x}} \delta L_{\mathbf{y}})$ for $\mathbf{x} \neq \mathbf{y}$, but resum all orders in local fluctuations $\sim \delta L_{\mathbf{x}}^n \delta L_{\mathbf{x}}^{*,m}$. The partition function factorizes $Z \approx z_{\text{r-mf}}^V$,
 \begin{align}
 	z_\text{r-mf} = e^{-d I(l,\bar{l},l,\bar{l})} \int \dif{W} \det Q_\text{stat}^\text{loc}(W) e^{2d I(L,L^*,l,\bar{l})}.
 \end{align} 
 Saddle points of the resummed free energy density $a^4f_{\text{r-mf}}(l, \bar{l}) = -\ln(z_\text{r-mf})/N_\tau$ are determined by
 \begin{align}
 \begin{pmatrix}
 	\frac{\partial}{\partial l}\\
 	\frac{\partial}{\partial \bar{l}}
 \end{pmatrix} I(l,\bar{l},l,\bar{l}) = \frac{2}{z_{\text{r-mf}}} \int \dif{W} \det Q_\text{stat}^\text{loc}(W) e^{2d I(L,L^*,l,\bar{l})} 
 \begin{pmatrix}
 	\frac{\partial}{\partial l}\\
 	\frac{\partial}{\partial \bar{l}}
 \end{pmatrix} I(L,L^*,l,\bar{l}),\label{eq:r-mf:Self_consistency_relation}
 \end{align}
 which implies for the self-consistent mean fields $l \neq \langle L \rangle_{\text{r-mf}}$ and $\bar{l} \neq \langle L^* \rangle_{\text{r-mf}}$ in general. Instead, equation \eqref{eq:r-mf:Self_consistency_relation} plays the role of the self-consistency relation. Neglecting terms at $\mathcal{O}(\delta L^2)$ from \eqref{eq:r-mf:Self_consistency_relation} restores the usual self-consistency relation. 
	\subsection{Classical approximation}
	In the two previous approaches the partition function \eqref{eq:pgStatic} was approximated by a single-site integral. However, solving the latter becomes numerically unstable when effective couplings become large. As seen in equations \eqref{eq:Definition_h1} and \eqref{eq:LO_kinetic_quark_eff_action}, this corresponds to $N_\tau \gg 1$, $\kappa \not\approx 0$, and $\mu_B\neq 0$, i.e. the low-temperature, finite-density regime with moderately heavy quarks. Due to the large effective couplings standard saddle-point methods are expected to be reliable. 
	
	We start by parameterizing Polyakov loops via two angles $\phi_1$ and $\phi_2$ by
	$L \rightarrow L(\phi_1,\phi_2) = e^{i\phi_1} + e^{i\phi_2} + e^{-i(\phi_1 + \phi_2)}$ and $ 
	L^* \rightarrow L(-\phi_1,-\phi_2)$.
	This transformation introduces a Jacobian that can be considered as an effective potential $V_{\text{eff}}$ \cite{Gross:1983ju},
	\begin{align}
	Z =& \frac{1}{(2\pi)^{2V}6^V} \int \left[\dif{\phi_{1,\mathbf{x}}}\right] \left[\dif{\phi_{2,\mathbf{x}}}\right] e^{-S^{\text{eff}} + V_{\text{eff}}} \\
	V_{\text{eff}} =& \sum_{\mathbf{x}} \ln\left(27 - 18|L_{\mathbf{x}}|^2 + 8\ReN(L_{\mathbf{x}}^3) - |L_{\mathbf{x}}|^4\right).
	\end{align}
	A leading-order saddle-point approximation for $\phi_1$ and $\phi_2$ around the saddle points $(\Phi_1,\Phi_2)$ of $\hat{S}^{\text{eff}} := S^{\text{eff}} - V_{\text{eff}}$ gives $Z\approx z_{\text{ca}}^V$, with
	\begin{align}
z_{\text{ca}}(\Phi_1,\Phi_2) :=& \frac{1}{6}\exp\left(-\frac{1}{V}\hat{S}^{\text{eff}}[L(\Phi_1,\Phi_2), L(-\Phi_1,-\Phi_2)]\right).
	\end{align}
	After defining $l := L(\Phi_1,\Phi_2)$ and $\bar{l} := L(-\Phi_1,-\Phi_2)$ one may locate the saddle-points of $a^4f_{\text{ca}}(l,\bar{l}) = -\ln(z_{\text{ca}})/N_\tau$ instead of directly determining $(\Phi_1,\Phi_2)$. For the expectation values of the Polyakov loops one can follow the same steps, giving $
		\langle L \rangle \approx \langle L \rangle_{\text{ca}} = l$ and $\langle L^* \rangle \approx\langle L^* \rangle_{\text{ca}} = \bar{l}$.
	\section{Mean field evaluation of the effective theories \label{sec:Evaluation}}
	We now apply the mean field approximations. To express results in physical units, we determine the lattice spacing using the Sommer parameter, $r_0$, in the pure gauge limit~\cite{Necco:2001xg}, 
	\begin{align}
		a(\beta) =& r_0\exp\left(-1.6804 - 1.7331(\beta-6) + 0.7849(\beta-6)^2 - 0.4428(\beta-6)^3\right), \label{eq:lattice_spacing}
	\end{align}
	for $5.7 \le \beta \le 6.92$. This remains valid as an approximation if quarks have finite but large masses. For the pion mass $am_\pi$ and baryon mass $am_B$ we use hopping-resummed expressions \cite{Smit_2002} with leading gauge corrections,
	\begin{align}
		am_\pi(\beta, \kappa) &= \mathrm{arccosh}\left[1 + \frac{(M^2 - 4)(M^2 - 1)}{2M^2-3}\right] - 24\kappa^2\frac{u}{1-u} + \dots\\
		am_B(\beta, \kappa) &= \ln\left[\frac{M^3(M^3-2)}{M^3-\frac{5}{4}}\right] - 18\kappa^2\frac{u}{1-u} + \dots,
	\end{align}
	where $M \equiv 1/(2\kappa)$. At any fixed $N_\tau$ the bare lattice parameters are related to the continuum temperature $T$ and pion mass $m_\pi$ by $am_\pi(\beta, \kappa)/a(\beta) = m_\pi $ and	$N_\tau am_\pi(\beta, \kappa) = m_\pi/T$.
	\subsection{Deconfinement transition\label{sec:eval_pure_gauge}}
			\begin{figure}[!t]
		\centering
		\begin{subfigure}[t]{0.48\textwidth}
			\includegraphics[width=\linewidth]{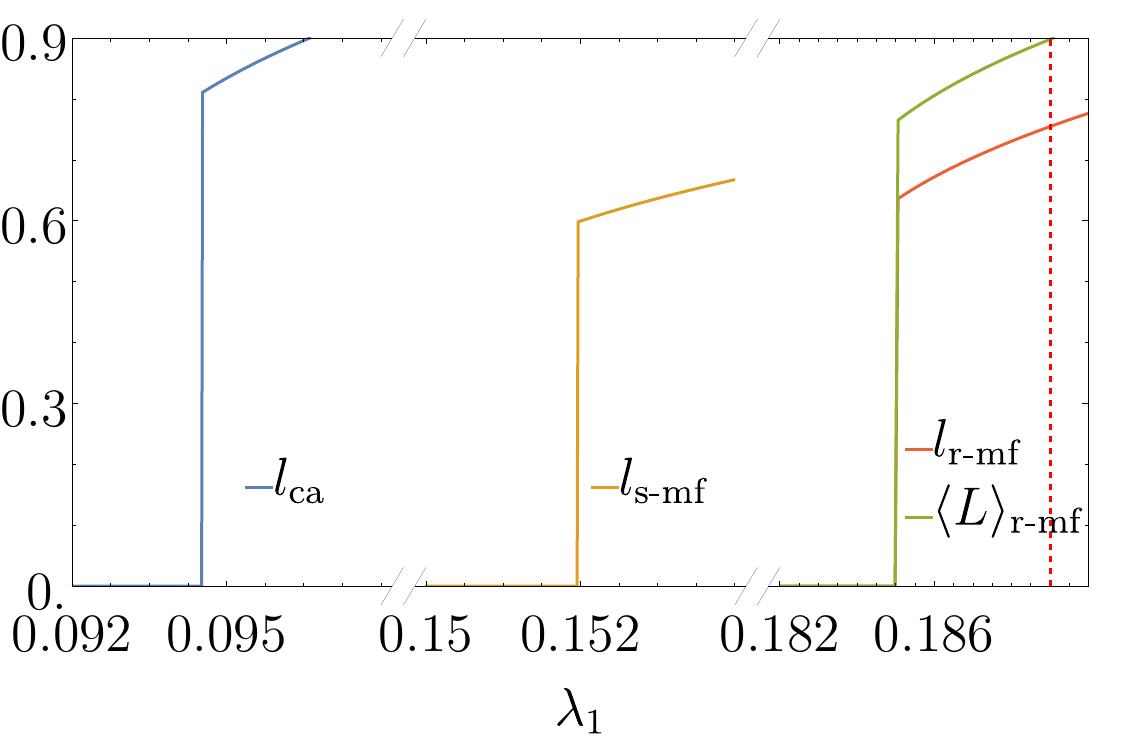}
		\end{subfigure}
		\quad
		\begin{subfigure}[t]{0.48\textwidth}
			\includegraphics[width=\linewidth]{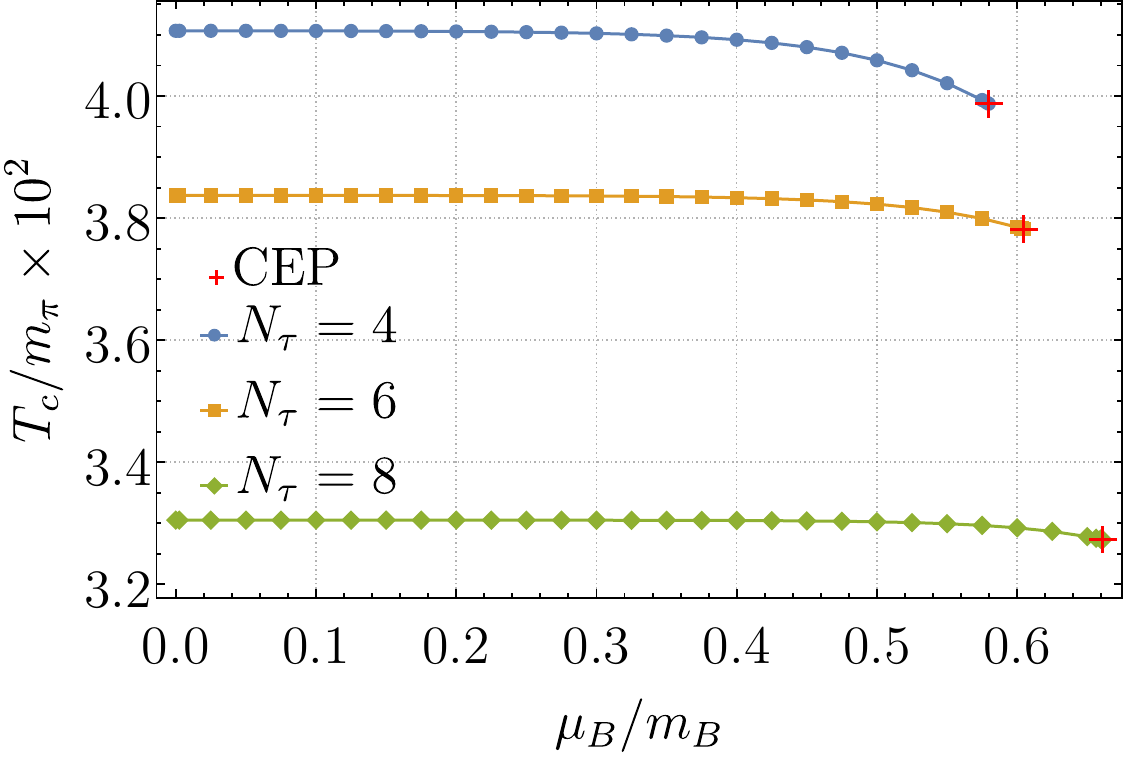}
		\end{subfigure}
		\caption{Comparison of the self-consistent mean fields in the pure gauge limit (left) and the phase diagram of the deconfinement transition at $\mu_B\neq 0$ (right).}
		\label{fig:resummedUnresummedAnddecon_finiteMuB_phaseDiagram}
	\end{figure}
	The mean field approaches are compared using the leading-order pure gauge effective action \eqref{eq:pure_gauge_effective_action} in figure \ref{fig:resummedUnresummedAnddecon_finiteMuB_phaseDiagram} (left), which shows the minimizing self-consistent mean field alongside a red dashed line indicating the critical coupling $\lambda_{1,c} \approx 0.1885$ obtained via series expansion~\cite{Kim:2019ykj}. Each approximation scheme shows the expected first-order transition, but the location of the transition differs drastically, reflecting the varying inclusion of local fluctuations. The ca approach, excluding any local fluctuations beyond leading order, yields $\lambda_{1,c} \approx 0.09$ (50\% relative error). The s-mf approach, accounting for fluctuations from the Haar-measure, improves this to $\lambda_{1,c} \approx 0.15$ (20\% relative error). The r-mf approach, incorporating all local fluctuations, achieves the most accurate result with $\lambda_{1,c} \approx 0.18505$ (3\% relative error). Additionally, the r-mf method reveals a split between the minimizing self-consistent mean field and the Polyakov loop expectation value, consistent with expectations from section \ref{sec:rmf}.
	
	We now switch to finite quark masses with $m_\pi = 15\text{GeV}$ and $\mu_B\neq 0$. Figure \ref{fig:resummedUnresummedAnddecon_finiteMuB_phaseDiagram} (right) shows the critical temperature for $N_f = 2$ and $N_\tau = 4, 6, 8$ as a function of $\mu_B/m_B$ obtained via the resummed mean field approach. For each $N_\tau$, the deconfinement transition is first-order and ends in a critical end point, in qualitative agreement with earlier effective theory studies \cite{Fromm2012}.
	\subsection{Nuclear liquid-gas transition \label{sec:eval_liquid_gas}}
					\begin{figure}[!t]\centering
					\begin{subfigure}[t]{0.48\textwidth}
						\centering
						\includegraphics[width=\linewidth,page=1]{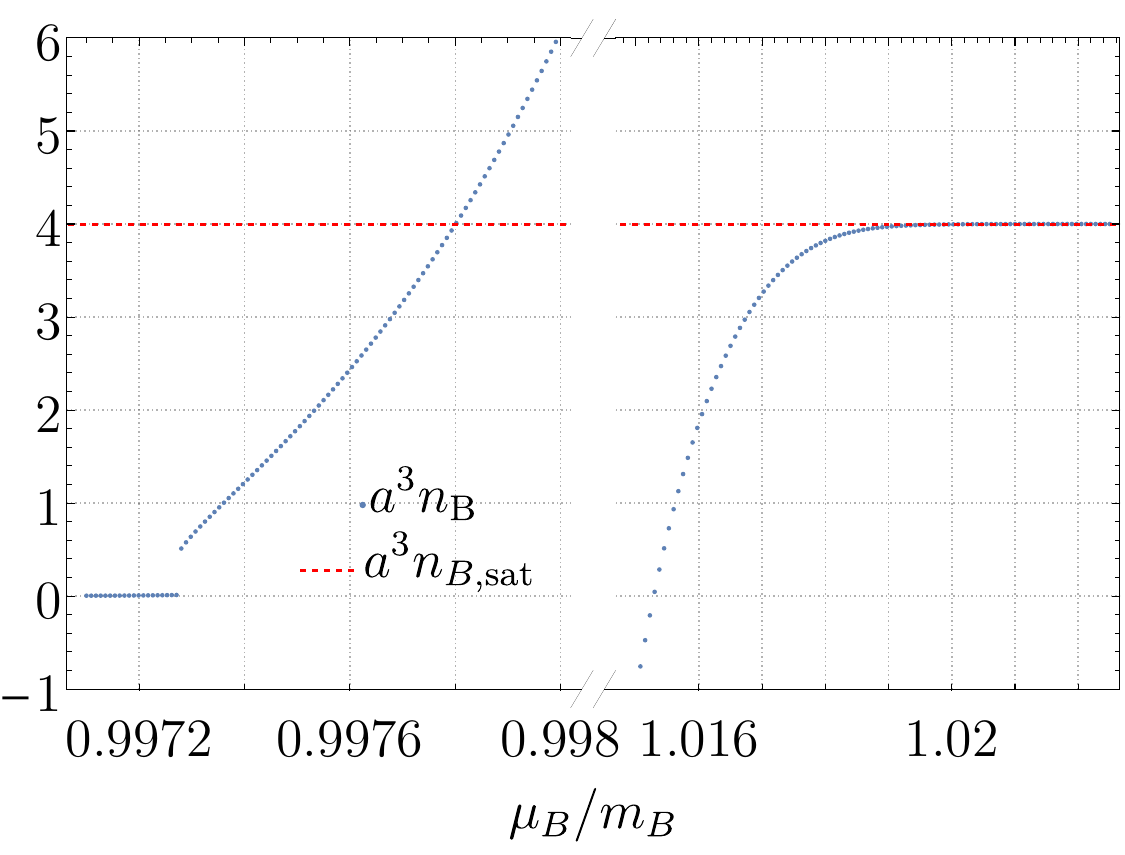}
					\end{subfigure}
					\quad
					\begin{subfigure}[t]{0.48\textwidth}
						\centering
						\includegraphics[width=\linewidth,page=1]{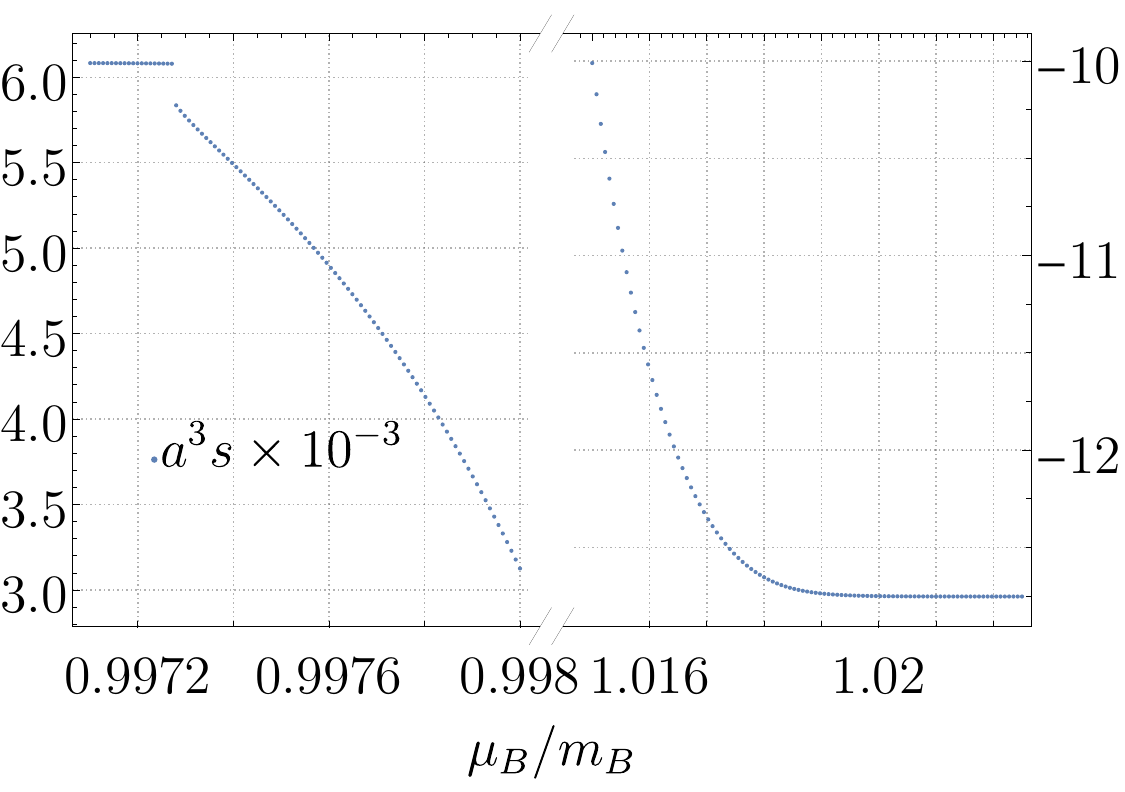}
					\end{subfigure}
					\caption{The baryon density $a^3n_B$ (left) and entropy density $a^3s$ (right) obtained via the ca approach around the nuclear liquid-gas transition and in the saturated regime. Also shown is the upper bound $a^3n_{B,\text{sat}} = 2N_f$ of $a^3n_B$ (red dashed line).}
					\label{fig:Finite_MuB_Densities}
				\end{figure} 
		\begin{figure}[!t]\centering
			\begin{subfigure}[t]{0.48\textwidth}
				\includegraphics[width=\linewidth,page=1]{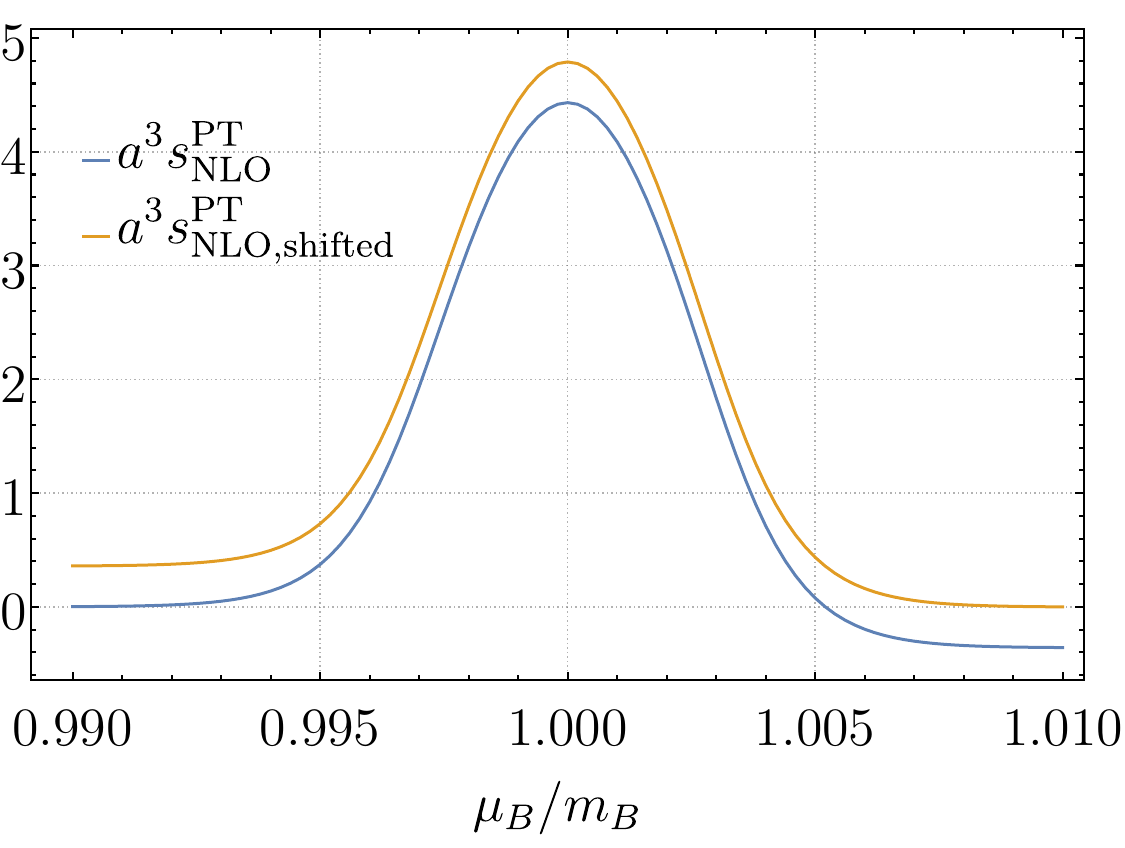}
			\end{subfigure}
			\quad
			\begin{subfigure}[t]{0.48\textwidth}
				\includegraphics[width=\linewidth,page=1]{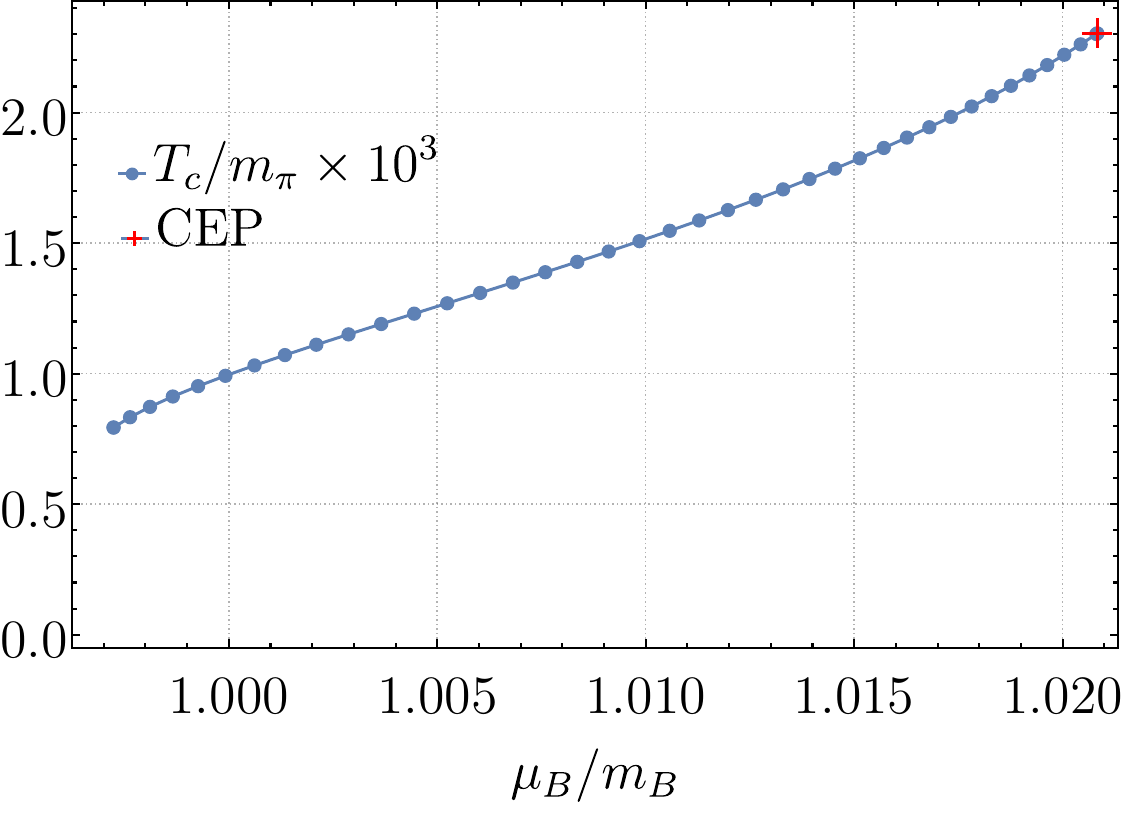}
			\end{subfigure}
			\caption{Perturbatively determined entropy densities (left) and nuclear liquid-gas transition line obtained via the ca approach (right).}
			\label{fig:Finite_MuB_entropy_density-perturbativeAndCrit_line}
	\end{figure}
We now consider more moderate pion masses, $m_\pi \approx 2.9$ GeV, at lower temperatures, $T/m_\pi \approx 7.9 \times 10^{-4}$, with $N_\tau = 500$ corresponding to $\kappa = 0.12$ and $\beta = 5.7$. At these parameters previous studies of the effective theories observed a first-order nuclear liquid-gas transition~\cite{Langelage:2014vpa}, although with an action incorporating fewer gauge corrections than the present work. To confirm this, we conducted $\mu_B$-direction scans using the ca approach.

Figure \ref{fig:Finite_MuB_Densities} (left) shows the baryon density $a^3n_B$ in the gaseous phase is nearly zero and jumps at the transition to a positive value, indicating the formation of a medium. Eventually $a^3 n_B$ surpasses $a^3 n_{B,\text{sat}} = 2N_f$, the saturation density implied by the Pauli-principle the lattice. As $\mu_B$ increases further, the self-consistent mean fields exceed the range $l, \bar{l} < 3$ in which we search for the $l, \bar{l}$\footnote{This is not strictly prohibited in the presence of the sign problem~\cite{Greensite:2012xv}}, until even larger values of $\mu_B$ are considered. Near lattice saturation the baryon density decreases to negative values before stabilizing at the expected saturation value, $n_{B,\text{sat}}$. This behavior of $a^3 n_B$ is unphysical as it violates convexity of the pressure, which may indicate a thermodynamical instability. An inhomogeneous phase could resolve this issue, but this behavior may also be a truncation artefact of the hopping expansion due to the large effective couplings. Violations of the Pauli-principle are possible because the truncated effective theory’s Boltzmann weight, $\exp\left(-S_{\text{eff}}\right)$, contains rational functions of $h_1 \approx e^{(\mu-m)/T}$ rather than the polynomial structure of the full quark determinant. As $\mu_B$ increases these rational dependencies cancel, leaving only contributions from the static determinant which correctly saturates at $2N_f$~\cite{Fromm2012}.

Figure \ref{fig:Finite_MuB_Densities} (right) presents the entropy density over the same $\mu_B$ intervals. The entropy jumps at the liquid-gas transition towards smaller values and decreases further with $\mu_B$, turning negative at lattice saturation. The jump towards smaller values is unexpected, because one expects the presence of a medium to increase the number of available states. As a qualitative benchmark for our ca approach we extend an earlier NLO pertubative analysis~\cite{Langelage:2014vpa} and use their results for the pressure $p$, energy density $e$ and baryon density $n_B$ to obtain the entropy density, shown in figure \ref{fig:Finite_MuB_entropy_density-perturbativeAndCrit_line} (left, blue line). As expected the perturbative entropy density vanishes in the vacuum and increases as the baryon onset is approached. In the saturated regime, it decreases and approaches a negative value, analytically determined as $-2dN_fN_\tau N_c\kappa^2$. This behavior aligns with our mean field results. Notably, the authors of~\cite{Langelage:2014vpa} performed their perturbative analysis to $O(\kappa^2)$ but set the scale using the leading-order pion mass expression, $am_\pi \approx -2\ln(2\kappa)$. Adjusting for this inconsistency by including the $O(\kappa^2)$ correction for $am_\pi$, the entropy density gains a shift of exactly $+2dN_fN_\tau N_c\kappa^2$, as shown in figure \ref{fig:Finite_MuB_entropy_density-perturbativeAndCrit_line} (left, orange line). Thus, inconsistent truncations for the effective action and the scale-setting can yield negative entropy density values.

The jump of the entropy density towards smaller values causes, by the Clausius-Clapeyron relation $\frac{\dif{T_c}}{\dif{\mu}} = -\frac{\Delta n}{\Delta s}$,
the critical temperature line to bend towards larger values of $\mu_B$ as the temperature is increased until it ends in a critical end point, as shown in figure \ref{fig:Finite_MuB_entropy_density-perturbativeAndCrit_line} (right). This is in contrast to expectations on liquid-gas transitions, and presumably
due to lattice and/or truncation artefacts.

	\section{Conclusions}
	This work applied mean field techniques to dimensionally reduced effective theories of LQCD, derived previously via strong coupling and hopping parameter expansions. 
	We introduced three variations of mean field approximations, each capturing different levels of fluctuations, and benchmarked them against the critical couplings in the pure gauge limit of the first order deconfinement transition. The least accurate method produced qualitative results (50\% error), while the most accurate approach achieved quantitative predictions within a few percent error. We then determined the critical endpoint of the deconfinement transition for heavy quarks at $\mu_B \neq 0$. In the low-temperature, finite-density regime with more moderate quark masses our mean field results confirmed a first-order nuclear liquid-gas transition. Near this transition, calculations of  thermodynamic observables revealed unphysical behavior in entropy and baryon number densities, likely due to truncations of the hopping expansion or inhomogeneous phase formation. We mapped out the first-order transition line and found it strongly influenced by unexpected entropy density behavior. Comparison with a perturbative analysis of the entropy showed qualitative agreement with our mean field results.
	
	Future work will address these limitations by deriving higher-order corrections to the effective actions and developing improved resummation methods, aiming to extend the theories’ validity to smaller quark masses. 
	\section*{Acknowledgments}
	We thank Jonas Scheunert for collaboration in the early stages of this project. The authors acknowledge support by the Deutsche Forschungsgemeinschaft (DFG, German Research Foundation) through the CRC-TR 211 'Strong-interaction matter under extreme conditions'- project number 315477589 - TRR 211 and by the State of Hesse within the Research Cluster ELEMENTS (Project ID 500/10.006).
\bibliographystyle{JHEP}
\bibliography{library}
\end{document}